\documentstyle[aps,epsfig]{revtex}

\begin{document}

\twocolumn[\hsize\textwidth\columnwidth\hsize\csname 
@twocolumnfalse\endcsname
\title{Negative Josephson coupling in the Kondo strong coupling limit}
\author{Oron Zachar}
\address{Laboratoire de Physique des Solides, U. Paris-Sud, 91405 Orsay , France.}
\date{15-Oct-98}
\maketitle

\widetext
\begin{abstract}
We Show that pair hopping through a Kondo singlet give rise to a negative 
Josephson coupling. Thus, our calculation supports the existance of 
staggered pair correlations in the strong coupling limit of a one dimensional Kondo lattice.
\end{abstract}

\vskip2pc]

\narrowtext

We use a simple strong coupling limit model, of an isotropic single channel
Kondo impurity, by which we show that BCS singlet-pair tunneling through a
Kondo singlet state results with a negative Josephson coupling between BCS
superconducting grains. For a lattice of such impurities (see figure-1), our
result implies a superconducting state with gap nodes due to the modulation,
or {\em staggering}, of the singlet pair correlation function. To avoid
possible misinterpretations, we stress that the node in the pair correlation
function due to negative Josephson coupling is a {\em node in the pair
center-of-mass motion}. It should not be confused with a node in the
relative pair state (such as p-wave or d-wave), which is usually thought 
\cite{Varma1987(Kondo)} to result from the residual repulsive interactions
between electrons on the screened impurity. We remark on the relation of our
result to the phase diagram of the 1D Kondo lattice.

In recent years, several theoretical approaches to the Kondo lattice problem
have yielded solutions with the interesting property of having a modulation
(or staggering) of the pairing order parameter at the Kondo impurities $%
x_{j} $; That is, the singlet pairing correlation function has the form, 
\begin{eqnarray}
\chi \left( x_{j}-x_{j^{^{\prime }}}\right) &=&\left\langle \hat{O}_{i}(x)%
\hat{O}_{i}^{\dagger }(x^{\prime })\right\rangle \\
&=&\left( -1\right) ^{\left( j-j^{\prime }\right) }\chi _{0}\left(
x_{j}-x_{j^{^{\prime }}}\right)  \nonumber
\end{eqnarray}
where $\chi _{0}$ is the usual correlation of the corresponding operator.
Coleman et al. \cite{Coleman-Miranda} have use an unusual mean field theory,
where a coherent three-body bound state of the impurity spins with the
conduction electrons leads to a state with the superconducting order
parameter fully staggered, i.e., the pair hopping amplitude from site to
site is negative. Zachar et al. \cite{zachar-KLL} have used a Bosonization
approach to a one dimensional Kondo lattice. They found staggered singlet
pairing correlations at a special exactly solvable ''Toulouse point'', which
correspond to a large value of the Kondo coupling coefficient. Heid et al. 
\cite{heid-Cox(stagger-Kondo)} have devised a new Ginzburg-Landau theory in
attempt to explain the superconducting transitions in $UPt_{3}$, $%
UPd_{2}Al_{3}$ and $UNi_{2}Al_{3}$. The distinctive feature of this theory 
\cite{heid-Cox(stagger-Kondo)} is a superconducting order parameter%
which resides {\it away} from the center of the Brillouin zone. Namely, they
have finite center-of-mass momentum or {\it staggered} pairing.

In all of the above mentioned theoretical approaches, the staggered
correlation were discussed in conjunction with a certain {\it two-channel}
Kondo\cite{2-ch-Kondo} feature leading to odd-$w$ (or odd-in-${\cal T}$)
pairing\cite{2-ch-Kondo,odd-w}. For an odd-$w$\ pair wave function it may be
shown
that, the pair transfer energy between two odd-$w$ slabs, i.e., Josephson
coupling, is negative--alternatively%
%
. Yet, the reverse is not necessarily true; having a staggered singlet
pairing does not entail odd-$w$ pairing. 

An undercurrent in all of the above approaches is a dominant Kondo singlet
resonance, and therefore, should intuitively be supported by a simple strong
coupling approximation. Indeed, Coleman et al. \cite{Coleman-effective-2ch}
have focussed on strong coupling models in which the bare single channel
Kondo coupling is turned into an effective multi-channel coupling due to
strong repulsive interactions within the conduction electron gas.

In this paper, we consider a simple strong coupling ''quantum chemistry''
type of model of a {\em single channel} Kondo impurity. The ground state on
the Kondo impurity is modeled as a single particle singlet bound-state, of
one conduction electron with the local impurity spin. This approach can be
considered as perturbations about the infinite Kondo coupling limit ($%
J_{K}/t\rightarrow \infty $). We investigate the problem of Josephson
coupling between two BCS superconducting grains through an intermediate
Kondo impurity site in a Kondo singlet state.

Our main result is that the Josephson coupling is {\em negative} across the
Kondo impurity, even though there is no odd-$w$ pairing. Therefore, we
demonstrate that staggered pairing correlations can be a consequence of the
Kondo singlet state and not only of odd-$w$ pairing.
\begin{figure}
\epsfxsize=3.0in 
\epsffile{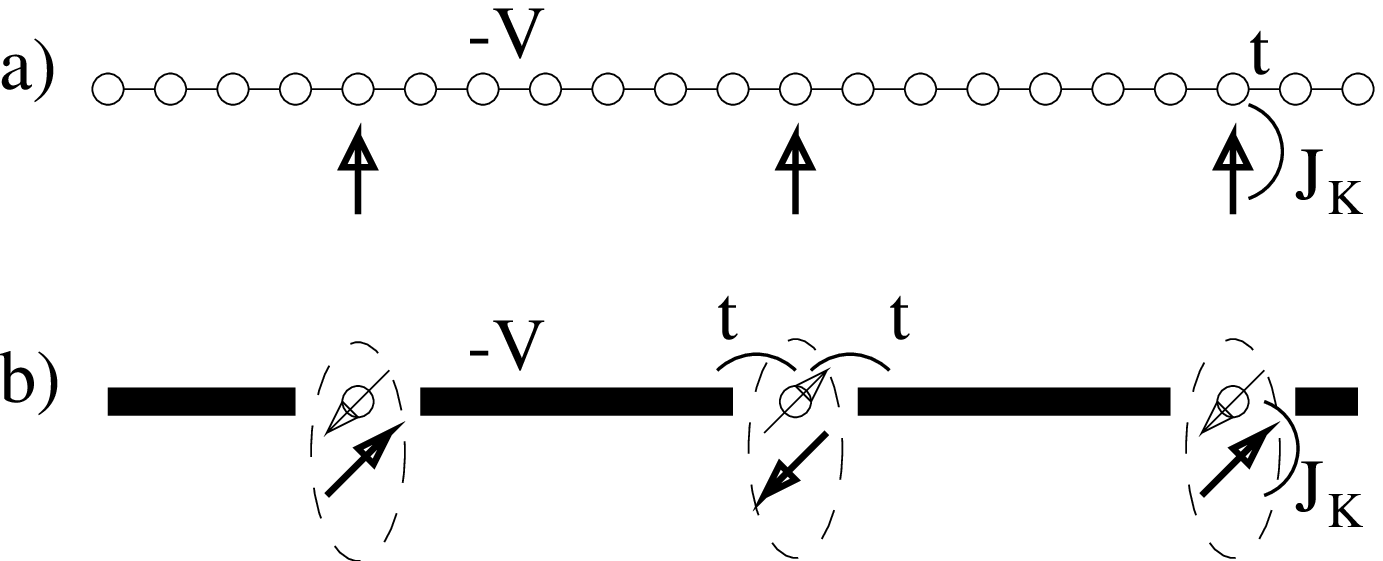}
\caption{}
\label{fig1}
\end{figure}
%
%
%

Consider a relatively dilute version of the one dimensional Kondo lattice
model, where the impurity lattice constant is much larger than the electron
lattice constant (see figure-1a). We work in the limit $J_{K}\gg t\gg \left|
-V\right| $, where $t$ is the hopping matrix element between nearest
neighbor electron lattice sites, $J_{K}$ is the Kondo coupling between one
local moment and one electron site (note: if the Kondo coupling would be to
more than one electron site then it will become a multi-channel Kondo
model), and $-V<0$ is an attractive interaction between the electrons (and
is of \ conventional origin, whose details are not important for our purpose
here). In this strong coupling limit, each Kondo impurity spin bounds a
single electron in a singlet state. The rest of the electrons in each
segment between impurities are going into a conventional BCS state, and thus
can be modeled as a BCS grain (see figure-1b).

The effective Josephson Hamiltonian is 
\begin{eqnarray}
H^{eff} &=&-2e^{2}\sum\limits_{ij}n_{i}\left[ C^{-1}\right]
_{ij}n_{j}+\sum\limits_{j}\mu _{j}n_{j}  \label{H-Josephson} \\
&&-\sum\limits_{ij}J_{ij}\cos \left( \theta _{i}-\theta _{j}+\frac{A_{ij}}{%
\Phi _{0}}\right)  \nonumber
\end{eqnarray}
where $2en_{j}$ is the charge on grain $j$, $C_{ij}$ is the capacitance, $%
A_{ij}$ is line integral of a vector potential from grain $i$ to grain $j$. $%
\left[ n_{j},e^{i\theta }\right] =\delta _{ij}e^{i\theta }$ .

In a Bose liquid, in the absence of a magnetic field, the kinetic energy is
unfrustrated in the sense that the ground state is real and nodeless. It is
therefore necessary that $J_{ij}>0$. Since ${\bf J}_{ij}{\bf =}\frac{J_{ij}}{%
\sin \left( \theta _{i}-\theta _{j}\right) }$ is the current across the
junction, $J_{ij}$ is a lattice version of the superfluid density $%
N_{s}\left( {\bf \ r}\right) $, (i.e. the proportionality constant between
the supercurrent and the velocity) 
\begin{equation}
J_{ij}\longleftrightarrow eN_{s}\left( {\bf r}\right) =\frac{{\bf J}%
_{s}\left( {\bf r}\right) }{{\bf v}\left( {\bf r}\right) }
\label{J-superfluid-density}
\end{equation}

Our aim here is to investigate the possibility of having {\em negative
Josephson coupling}, (or negative superfluid density) $J_{ij}<0$,%
due to the presence of Kondo impurities in {\em the strong coupling limit}.

We investigate the simplest model, in which there are only two
superconducting grains. In the usual case, in which {\em direct}
single-electron tunneling between grains is responsible for Josephson
coupling, $J_{12}$ is guaranteed to be positive in the absence of spin-orbit
coupling. However, the situation is much richer if we consider the case in
which the tunneling is indirect, through a localized state\cite
{Spivak-Kivelson}. We consider the effect of a Kondo impurity situated
between the two superconducting grains. Above the Kondo temperature, the
effect of spin-flip scattering \cite{Josephson-SpinFlip} of the tunneling
electrons by the impurity spin, (which is equivalent to the effect of
spin-orbit coupling), can produce $J_{12}<0$ . Typically though, a
superconducting transition happens well below the Kondo temperature (e.g.,
in heavy fermion materials).

To investigate the effect of an {\em intermediate} coherent Kondo state on
the Josephson coupling, we examine a strong coupling model. In that case,
the ground state of the system has a single electron bound in a singlet
state with the impurity spin. We supplement the impurity Kondo interaction
with a residual repulsive on-site interaction, $U_{i}$, in order to
investigate the effect of such a modification, and also in order to keep
track of the double occupied impurity states. For simplicity, we neglect
direct hopping between the two superconducting grains. In the context of one
dimensional systems, such as the Kondo array (see figure), this neglect is
justified since the direct Josephson coupling between grains is
exponentially small compared with the coupling to the impurity (which is
situated in the middle between the left and right ''grains'').

The impurity spectrum is 
\[
\begin{tabular}{|l|l|}
\hline
Energy & eigenstates \\ \hline
$E_{2}=+\frac{1}{4}J_{K}$ & $\left| \Uparrow \uparrow \right\rangle ,\left|
\Downarrow \downarrow \right\rangle ,\left| T_{0}\right\rangle \equiv \frac{1%
}{\sqrt{2}}\left( \left| \Uparrow \downarrow \right\rangle +\left|
\Downarrow \uparrow \right\rangle \right) $ \\ \hline
$E_{1}=0$ & $\left| \Uparrow \right\rangle ,\left| \Downarrow \right\rangle $
\\ \hline
$E_{1}^{^{\prime }}=0+U_{i}$ & $\left| \Uparrow \uparrow \downarrow
\right\rangle ,\left| \Downarrow \uparrow \downarrow \right\rangle $ \\ 
\hline
$E_{0}=-\frac{3}{4}J_{K}$ & $\left| S_{i}\right\rangle \equiv \frac{1}{\sqrt{%
2}}\left( \left| \Uparrow \downarrow \right\rangle -\left| \Downarrow
\uparrow \right\rangle \right) $ \\ \hline
\end{tabular}
\]
where $\Uparrow $ is the impurity spin and $\uparrow $ is a conduction
electron spin. The Hamiltonian consists of the two superconducting grains, $%
H_{1}\left[ a\right] $ and $H_{2}\left[ b\right] $, the Kondo impurity site
interactions ($J_{K}$ and $U_{i}$), and hopping interaction, $T$, between
the impurity and each grain. 
\begin{eqnarray}
H &=&H_{0}+T \\
H_{0} &=&H_{1}\left( a\right) +H_{2}\left( b\right) +\left[ J_{K}{\bf \tau
\cdot S}+U_{i}\left( n_{0}\right) ^{2}\right]  \label{H0} \\
T &=&\sum\limits_{ks}\left[ T_{k}\left( b_{ks}^{\dagger
}c_{0s}+c_{0s}^{\dagger }a_{ks}\right) +{\rm h.c.}\right]  \label{T-hop}
\end{eqnarray}

The lowest order contribution to the Josephson coupling, through
intermediate excitations of the impurity state, is forth order in the
hopping interaction, 
\[
J^{\left( 4\right) }=T\frac{1}{w-H_{0}+i\varepsilon }T\frac{1}{%
w-H_{0}+i\varepsilon }T\frac{1}{w-H_{0}+i\varepsilon }T 
\]
\begin{eqnarray}
J\cos ( &&\theta _{1}-\theta _{2})  \nonumber \\
&=&-\left\langle S_{i}\right| \left\langle \psi _{1}\left( a\right) \psi
_{2}\left( b\right) \right| \hat{J}^{\left( 4\right) }\left| \psi _{1}\left(
a\right) \psi _{2}\left( b\right) \right\rangle \left| S_{i}\right\rangle 
\nonumber \\
&=&+\left\langle S_{i}\right| \left\langle \psi _{1}\psi _{2}\right| T\frac{1%
}{H_{0}}T\frac{1}{H_{0}}T\frac{1}{H_{0}}T\left| \psi _{1}\psi
_{2}\right\rangle \left| S_{i}\right\rangle .
\end{eqnarray}
Where $\left| S_{i}\right\rangle =\frac{1}{\sqrt{2}}\left( \left| \Uparrow
\downarrow \right\rangle -\left| \Downarrow \uparrow \right\rangle \right) $
is the impurity singlet ground state, and the $\psi _{j}$ are BCS ground
state wave functions on grains $1$ and $2$ respectively, 
\begin{equation}
\left| \psi _{j}\right\rangle =\prod\limits_{q}\left( u_{q}+e^{i\theta
_{j}}v_{q}c_{jq\uparrow }^{\dagger }c_{j-q\downarrow }^{\dagger }\right)
\left| 0\right\rangle .  \label{BCS-state}
\end{equation}

It is necessary to define a case convention (i.e., the order of spin up/down
creation operators). We chose to put up-spin to the left of down-spin
operators in the BCS wave-function (\ref{BCS-state}). Thus the proper
ordering of the operators that transfer a pair from the left grain to the
right grain is $\left\langle \psi _{1}\psi _{2}\right| \left( b_{q\uparrow
}^{\dagger }b_{-q\downarrow }^{\dagger }a_{-k\downarrow }a_{k\uparrow
}\right) \left| \psi _{1}\psi _{2}\right\rangle $. Direct hoping between
grains preserves this order, and results with positive Josephson coupling, $%
J>0$. As we shall see, in tunneling processes through the Kondo impurity
site, the ordering can get permuted, leading to a $\left( -\right) $ sign.
This effect can be traced to the single electron occupancy of the impurity
ground state \cite{Spivak-Kivelson}.

To fourth order in the hopping interaction, there are $12$ different
processes that transfer a pair from the left grain to the right grain (and
similarly $12$ Hermitian conjugate processes that transfer from the right to
the left). The $12$ processes can be grouped into six pairs of processes
with identical amplitudes (one of which has the intermediate impurity spin
up, $\Uparrow $, and the other process has the intermediate impurity spin
down, $\Downarrow $ ). The contribution of each and {\em every one of the
processes is negative} , leading to an effective negative Josephson coupling
across the Kondo impurity. This is our main result. 
\begin{eqnarray}
J^{eff} &=&\left( J_{1}+J_{2}+J_{3}+J_{4}+J_{5}+J_{6}\right)  \label{J-eff}
\\
&=&-{\cal M}\sum\limits_{kq}\left| T_{k}T_{q}\right| ^{2}\left( v_{q}^{\ast
}u_{k}^{\ast }v_{k}u_{q}\right)
\end{eqnarray}
\begin{equation}
{\cal M}=\left[ 
\begin{array}{c}
+\frac{1}{\left( \varepsilon _{k}+\frac{3}{4}J_{K}\right) \left(
J_{K}+\varepsilon _{k}+\varepsilon _{q}\right) \left( \frac{3}{4}%
J_{K}+\varepsilon _{q}\right) } \\ 
+\frac{1}{\left( \varepsilon _{k}+\frac{3}{4}J_{K}+U_{i}\right) \left(
J_{K}+\varepsilon _{k}+\varepsilon _{q}\right) \left( \frac{3}{4}%
J_{K}+\varepsilon _{q}\right) } \\ 
+\frac{1}{2}\frac{1}{\varepsilon _{k}+\frac{3}{4}J_{K}+U_{i}}\left( \frac{1}{%
\varepsilon _{k}+\varepsilon _{q}+J_{K}}+\frac{1}{\varepsilon
_{k}+\varepsilon _{q}}\right) \frac{1}{\frac{3}{4}J_{K}+\varepsilon _{q}} \\ 
+\frac{1}{\left( \varepsilon _{k}+\frac{3}{4}J_{K}\right) \left(
J_{K}+\varepsilon _{k}+\varepsilon _{q}\right) \left( \frac{3}{4}%
J_{K}+U_{i}+\varepsilon _{k}\right) } \\ 
+\frac{1}{\left( \varepsilon _{q}+\frac{3}{4}J_{K}+U_{i}\right) \left(
J_{K}+\varepsilon _{k}+\varepsilon _{q}\right) \left( \frac{3}{4}%
J_{K}+U_{i}+\varepsilon _{k}\right) } \\ 
+\frac{1}{2}\frac{1}{\varepsilon _{k}+\frac{3}{4}J_{K}}\left( \frac{1}{%
\varepsilon _{k}+\varepsilon _{q}+J_{K}}+\frac{1}{\varepsilon
_{k}+\varepsilon _{q}}\right) \frac{1}{\frac{3}{4}J_{K}+U_{i}+\varepsilon
_{k}}
\end{array}
\right]  \label{J-amplitude}
\end{equation}
(Note that, by definition, all the denominators must be positive since we
are considering virtual excitations relative to the unperturbed
ground-state). The first process (see table (1A) in the appendix), is
identical to the negative Josephson coupling derivation in reference \cite
{Spivak-Kivelson}, which inspired our calculation here. That process would
dominate in the limit of large repulsion, $U_{i}\rightarrow \infty $, on the
impurity site (as indeed was the case considered in reference \cite
{Spivak-Kivelson}). In contrast, we consider the limit where $U_{i}\ll J_{K}$
can take any negligible small value. In that case, all the $12$ processes
have contribution of the same order of magnitude. In the appendix we list
the contributing processes, and show in two explicit examples how the
negative sign comes about. 

In conclusion, for antiferromagnetic Kondo interactions, $J_{K}/t\gg 1$, 
{\em all } the contributions to the Josephson coupling, from intermediate
hopping through the impurity states, $%
J^{eff}=J_{1}+J_{2}+J_{3}+J_{4}+J_{5}+J_{6}$, are negative. The {\em residual%
} repulsive interactions on the impurity site, $U_{i}$, have only a
quantitative effect.

In a continuum limit, our result can be associated with a recent one
dimensional field theoretic derivation by Salkola et al.\cite
{Salkola-SCkinks}; that a local electronic state of charge-$1e$ spin-$0$
leads to a $\pi -$kink in the pairing order parameter.

For a periodic one dimensional Kondo lattice of such impurities (figure-1),
in the {\em strong coupling limit}, the pair correlation function is {\em %
staggered, due to negative Josephson coupling across impurity sites}, $\chi
\left( x_{j}-x_{j^{^{\prime }}}\right) =\left( -1\right) ^{\left(
j-j^{\prime }\right) }\chi _{0}\left( x_{j}-x_{j^{^{\prime }}}\right) $.
Where $\chi _{0}$ has power law decay (as in a 1D negative $U$ Hubbard
model).

The form of the staggered BCS pairing, written above, is identical to the
one which we found previously in a Bosonization approach to the 1D Kondo
lattice \cite{zachar-KLL,Z-staggered-liquids}. That Bosonization calculation
was valid in a limited region of parameter space. The fact that we find the
same staggered pairing state in a strong coupling limit (in this paper)
leads us to propose that a phase of staggered BCS pairing correlations
extends over a wide region of parameter space in the phase diagram of the 1D
Kondo lattice model. 
Note that the staggered pairing order, found in this paper, is in the strong
coupling limit of a {\em single channel} Kondo lattice, independently of $%
odd-w$ pairing and two-channel physics.

{\bf Acknowledgments:} I thank S. Kivelson for innumerable stimulating
discussion about the 1D Kondo lattice, and P. Coleman for helpful comments
on $odd-w$ pairing. This work was supported by the TMR Fellowship
\#ERB4001GT97294.

\subsection{Appendix: Calculation of Josephson coupling terms}

Below, I give a graphical depiction of the 6 processes $\left( 1A\rightarrow
6A\right) $ which transfer a pair from $grain-S_{1}$ to $grain-S_{2}$ , with
an intermediate impurity $spin-\Downarrow $ state. (In the initial and final
ground states, the impurity site is in a singlet bound state; $\left|
S_{i}\right\rangle \equiv \frac{1}{\sqrt{2}}\left( \left| \Uparrow
\downarrow \right\rangle -\left| \Downarrow \uparrow \right\rangle \right) $%
). The left column is the intermediate state energy. Since the initial and
final states are the same for all the processes, we indicate them only in
the table of the first process: 
\begin{eqnarray*}
&&\frame{$
\begin{array}{cccc}
\left( 1A\right) {\bf energy} & \left( S_{1}\right) & {\bf impurity} & 
\left( S_{2}\right) \\ 
0: & \left( \uparrow \downarrow \right) & \left| S_{i}\right\rangle & \left(
{}\right) \\ 
\frac{3}{4}J_{K}+\varepsilon _{q}: & \left( \uparrow \downarrow \right) & 
\left( \Downarrow \right) & \left( \uparrow _{q}\right) \\ 
J_{K}+\varepsilon _{k}+\varepsilon _{q}: & \left( \uparrow _{k}\right) & 
\left( \Downarrow \downarrow \right) & \left( \uparrow _{q}\right) \\ 
\varepsilon _{k}+\frac{3}{4}J_{K}: & \left( \uparrow _{k}\right) & \left(
\Downarrow \right) & \left( \downarrow \uparrow \right) \\ 
0: & \left( {}\right) & \left( \Downarrow \uparrow \right) & \left(
\downarrow \uparrow \right) \\ 
0: & \left( {}\right) & \left| S_{i}\right\rangle & \left( \uparrow
\downarrow \right)
\end{array}
$} \\
&&\frame{$
\begin{array}{cccc}
\left( 2A\right) {\bf energy} & \left( {\bf S}_{1}\right) & {\bf impurity} & 
\left( {\bf S}_{2}\right) \\ 
\frac{3}{4}J_{K}+\varepsilon _{q}: & \left( \uparrow \downarrow \right) & 
\left( \Downarrow \right) & \left( \uparrow _{q}\right) \\ 
J_{K}+\varepsilon _{k}+\varepsilon _{q}: & \left( \uparrow _{k}\right) & 
\left( \Downarrow \downarrow \right) & \left( \uparrow _{q}\right) \\ 
\varepsilon _{q}+\frac{3}{4}J_{K}+U_{i}: & \left( {}\right) & \left(
\Downarrow \uparrow \downarrow \right) & \left( \uparrow _{q}\right)
\end{array}
$} \\
&&\frame{$
\begin{array}{cccc}
\left( 3A\right) {\bf energy} & \left( {\bf S}_{1}\right) & {\bf impurity} & 
\left( {\bf S}_{2}\right) \\ 
\frac{3}{4}J_{K}+\varepsilon _{q}: & \left( \uparrow \downarrow \right) & 
\left( \Downarrow \right) & \left( \uparrow _{q}\right) \\ 
\left( 
\begin{array}{c}
\frac{1}{\varepsilon _{k}+\varepsilon _{q}+J_{K}} \\ 
+\frac{1}{\varepsilon _{k}+\varepsilon _{q}}
\end{array}
\right) : & \left( \downarrow _{-k}\right) & \left( \Downarrow \uparrow
\right) & \left( \uparrow _{q}\right) \\ 
\varepsilon _{q}+\frac{3}{4}J_{K}+U_{i}: & \left( {}\right) & \left(
\Downarrow \uparrow \downarrow \right) & \left( \uparrow _{q}\right)
\end{array}
$} \\
&&\frame{$
\begin{array}{cccc}
(4A){\bf energy} & \left( S_{1}\right) & {\bf impurity} & \left( S_{2}\right)
\\ 
\frac{3}{4}J_{K}+U_{i}+\varepsilon _{k}: & \left( \uparrow _{k}\right) & 
\left( \Downarrow \uparrow \downarrow \right) & \left( {}\right) \\ 
\varepsilon _{k}+\varepsilon _{q}+J_{K}: & \left( \uparrow _{k}\right) & 
\left( \Downarrow \downarrow \right) & \left( \uparrow _{q}\right) \\ 
\varepsilon _{k}+\frac{3}{4}J_{K}: & \left( \uparrow _{k}\right) & \left(
\Downarrow \right) & \left( \downarrow \uparrow \right)
\end{array}
$} \\
&&\frame{$
\begin{array}{cccc}
(5A){\bf energy} & \left( S_{1}\right) & {\bf impurity} & \left( S_{2}\right)
\\ 
\frac{3}{4}J_{K}+U_{i}+\varepsilon _{k}: & \left( \uparrow _{k}\right) & 
\left( \Downarrow \uparrow \downarrow \right) & \left( {}\right) \\ 
\varepsilon _{k}+\varepsilon _{q}+J_{K}: & \left( \uparrow _{k}\right) & 
\left( \Downarrow \downarrow \right) & \left( \uparrow _{q}\right) \\ 
\varepsilon _{q}+\frac{3}{4}J_{K}+U_{i}: & \left( {}\right) & \left(
\Downarrow \uparrow \downarrow \right) & \left( \uparrow _{q}\right)
\end{array}
$} \\
&&\frame{$
\begin{array}{cccc}
\left( 6A\right) {\bf energy} & \left( {\bf S}_{1}\right) & {\bf impurity} & 
\left( {\bf S}_{2}\right) \\ 
\frac{3}{4}J_{K}+U_{i}+\varepsilon _{k}: & \left( \uparrow _{k}\right) & 
\left( \Downarrow \uparrow \downarrow \right) & \left( {}\right) \\ 
\left( 
\begin{array}{c}
\frac{1}{\varepsilon _{k}+\varepsilon _{q}+J_{K}} \\ 
+\frac{1}{\varepsilon _{k}+\varepsilon _{q}}
\end{array}
\right) : & \left( \uparrow _{k}\right) & \left( \Downarrow \uparrow \right)
& \left( \downarrow _{-q}\right) \\ 
\varepsilon _{k}+\frac{3}{4}J_{K}: & \left( \uparrow _{k}\right) & \left(
\Downarrow \right) & \left( \uparrow \downarrow \right)
\end{array}
$}
\end{eqnarray*}

I demonstrate the calculation in two cases; In case 1A, the origin of the $%
\left( -\right) $ sign is from the operator ordering in the BCS wave
function of the grains (i.e., to the order of operation $\left\langle \psi
_{1}\psi _{2}\right| \left( b_{q\uparrow }^{\dagger }b_{-q\downarrow
}^{\dagger }a_{-k\downarrow }a_{k\uparrow }\right) \left| \psi _{1}\psi
_{2}\right\rangle $). In case 4A, the $\left( -\right) $ sign results from
ordering of the electron operators on the impurity site, $\left( 1-n_{0\bar{%
\sigma}}\right) n_{0\sigma }=\left( c_{0\bar{\sigma}}c_{0\bar{\sigma}%
}^{\dagger }c_{0\sigma }^{\dagger }c_{0\sigma }\right) $. In all of the 12
pair transfer processes, one of these fermion operator orderings takes
place, leading to the negative sign.

Case-1A:

\begin{eqnarray*}
&&J\cos \left( \theta _{1}-\theta _{2}\right) \\
&=&+\left\langle S_{i}\right| \left\langle \psi _{1}\psi _{2}\right|
\sum\limits_{kqk^{^{\prime }}q^{^{\prime }}}T_{k^{^{\prime }}}\left(
c_{0\uparrow }^{\dagger }a_{k^{^{\prime }}\uparrow }\right) \times \\
&&\frac{T_{q^{^{\prime }}}\left( b_{q^{^{\prime }}\downarrow }^{\dagger
}c_{0\downarrow }\right) }{\left( \varepsilon _{k}+\frac{3}{4}J_{K}\right) }%
\frac{T_{-k}\left( c_{0\downarrow }^{\dagger }a_{-k\downarrow }\right) }{%
\left( J_{K}+\varepsilon _{k}+\varepsilon _{q}\right) }\frac{T_{q}\left(
b_{q\uparrow }^{\dagger }c_{0\uparrow }\right) }{\left( \frac{3}{4}%
J_{K}+\varepsilon _{q}\right) }\left| \psi _{1}\psi _{2}\right\rangle \left|
S_{i}\right\rangle \\
&=&+\sum\limits_{kq}\frac{1}{2}\left| T_{k}T_{q}\right| ^{2}\left\langle
\Downarrow \uparrow \right| \left( c_{0\downarrow }c_{0\downarrow }^{\dagger
}c_{0\uparrow }^{\dagger }c_{0\uparrow }\right) \left| \Downarrow \uparrow
\right\rangle \times \\
&&\frac{\left\langle \psi _{1}\psi _{2}\right| \left[ -\left( b_{q\uparrow
}^{\dagger }b_{-q\downarrow }^{\dagger }a_{-k\downarrow }a_{k\uparrow
}\right) \right] \left| \psi _{1}\psi _{2}\right\rangle }{\left( \varepsilon
_{k}+\frac{3}{4}J_{K}\right) \left( J_{K}+\varepsilon _{k}+\varepsilon
_{q}\right) \left( \frac{3}{4}J_{K}+\varepsilon _{q}\right) } \\
&=&-\frac{1}{2}\sum\limits_{kq}\frac{\left| T_{k}T_{q}\right| ^{2}\left(
e^{-i\theta _{2}}v_{q}^{*}u_{k}^{*}\right) \left( e^{i\theta
_{1}}v_{k}u_{q}\right) }{\left( \varepsilon _{k}+\frac{3}{4}J_{K}\right)
\left( J_{K}+\varepsilon _{k}+\varepsilon _{q}\right) \left( \frac{3}{4}%
J_{K}+\varepsilon _{q}\right) }.
\end{eqnarray*}
Thus, 
\[
J_{1}=-\sum\limits_{kq}\frac{\left| T_{k}T_{q}\right| ^{2}\left(
v_{q}^{*}u_{k}^{*}v_{k}u_{q}\right) }{\left( \varepsilon _{k}+\frac{3}{4}%
J_{K}\right) \left( J_{K}+\varepsilon _{k}+\varepsilon _{q}\right) \left( 
\frac{3}{4}J_{K}+\varepsilon _{q}\right) }. 
\]

The factor $\left| T_{k}T_{q}\right| ^{2}\left(
v_{q}^{*}u_{k}^{*}v_{k}u_{q}\right) $ is common to all the terms. All we
need to do is figure out the sign of the permutation to canonical order of
operators and the energy denominator of each possible tunneling sequence.

Case-4A:

\begin{eqnarray*}
J &=&+\sum\limits_{kq}\frac{1}{2}\left| T_{k}T_{q}\right| ^{2}\left\langle
\Downarrow \uparrow \right| \left\langle \psi _{1}\psi _{2}\right| \\
&&\frac{\left( c_{0\uparrow }^{\dagger }a_{k\uparrow }\right) \left(
b_{-q\downarrow }^{\dagger }c_{0\downarrow }\right) \left( b_{q\uparrow
}^{\dagger }c_{0\uparrow }\right) \left( c_{0\downarrow }^{\dagger
}a_{-k\downarrow }\right) \left| \psi _{1}\psi _{2}\right\rangle \left|
\Downarrow \uparrow \right\rangle }{\left( \varepsilon _{k}+\frac{3}{4}%
J_{K}\right) \left( J_{K}+\varepsilon _{k}+\varepsilon _{q}\right) \left( 
\frac{3}{4}J_{K}+U+\varepsilon _{k}\right) } \\
&=&+\sum\limits_{kq}\frac{1}{2}\left| T_{k}T_{q}\right| ^{2}\left\langle
\Downarrow \uparrow \right| \left[ -\left( c_{0\downarrow }c_{0\downarrow
}^{\dagger }c_{0\uparrow }^{\dagger }c_{0\uparrow }\right) \right] \left|
\Downarrow \uparrow \right\rangle \\
&&\times \frac{\left\langle \psi _{1}\psi _{2}\right| \left( b_{q\uparrow
}^{\dagger }b_{-q\downarrow }^{\dagger }a_{-k\downarrow }a_{k\uparrow
}\right) \left| \psi _{1}\psi _{2}\right\rangle }{\left( \varepsilon _{k}+%
\frac{3}{4}J_{K}\right) \left( J_{K}+\varepsilon _{k}+\varepsilon
_{q}\right) \left( \frac{3}{4}J_{K}+U+\varepsilon _{k}\right) }.
\end{eqnarray*}
Thus 
\[
J_{4}=-\sum\limits_{kq}\frac{\left| T_{k}T_{q}\right| ^{2}\left(
v_{q}^{*}u_{k}^{*}v_{k}u_{q}\right) }{\left( \varepsilon _{k}+\frac{3}{4}%
J_{K}\right) \left( J_{K}+\varepsilon _{k}+\varepsilon _{q}\right) \left( 
\frac{3}{4}J_{K}+U+\varepsilon _{k}\right) }. 
\]


\end{document}